\RequirePackage{rotating}
\documentclass[traditabstract,twocolumn]{aa} 

\usepackage{epsfig,graphicx,natbib}
\usepackage[normalem]{ulem}
\usepackage{amssymb}
\usepackage{subfig}
\usepackage[flushmargin]{footmisc}
\usepackage{rotating}

\def\nH2{\hbox{$n_\mathrm{H_2}$}}
\def\kms{\hbox{km\,s$^{-1}$}}
\def\msun{\hbox{M$_\odot$}}
\def\PKS1830{\hbox{PKS\,1830$-$211}}
\def\B0218{\hbox{B\,0218+357}}
\def\cm-2{\hbox{cm$^{-2}$}}
\def\cm-3{\hbox{cm$^{-3}$}}
\def\H2{\hbox{H$_2$}}
\def\Rab{\hbox{$R_{A/B}$}}

\begin{document}

\title{Isotopic ratios at $z$=0.68 from molecular absorption lines toward B\,0218+357} 

\author{S. H. J. Wallstr\"om \inst{1}
\and S. Muller \inst{1} 
\and M. Gu\'elin \inst{2,3}
}

\institute{
Department of Earth and Space Sciences, Chalmers University of Technology, Onsala Space Observatory, SE--43992 Onsala, Sweden
\and Institut de Radio Astronomie Millim\'etrique, 300 rue de la piscine, F--38406 St Martin d'H\`eres, France
\and Ecole Normale Sup\'erieure/LERMA, 24 rue Lohmond, F--75005 Paris, France
}

\date {Received / Accepted }

\titlerunning{Isotopic ratios toward B\,0218+357}
\authorrunning{Wallstr\"om et al.}

\abstract{Isotopic ratios of heavy elements are a key signature of the nucleosynthesis processes in stellar interiors. The contribution of successive generations of stars to the metal enrichment of the Universe is imprinted on the evolution of isotopic ratios over time. We investigate the isotopic ratios of carbon, nitrogen, oxygen, and sulfur through millimeter molecular absorption lines arising in the $z$=0.68 absorber toward the blazar B\,0218+357. We find that these ratios differ from those observed in the Galactic interstellar medium, but are remarkably close to those in the only other source at intermediate redshift for which isotopic ratios have been measured to date, the $z$=0.89 absorber in front of PKS\,1830$-$211. The isotopic ratios in these two absorbers should reflect enrichment mostly from massive stars, and they are indeed close to the values observed toward local starburst galaxies. Our measurements set constraints on nucleosynthesis and chemical evolution models.}

\keywords{quasars: absorption lines -- quasars: individual: B\,0218+357 -- galaxies: ISM -- galaxies: abundances -- ISM: molecules -- radio lines: galaxies}

\maketitle

\section{Introduction}

Isotopic ratios provide a powerful way to study the origin of elements and the chemical evolution of the Universe. They can be measured to extremely high precision in the lab, using mass spectrometry. However, this is limited to physical samples such as presolar grains in meteorites and solar wind particles. For any remote astronomical sources, we must rely on spectroscopic analysis, with limits on angular resolution, sensitivity, and knowledge of the physical conditions.
Here molecular lines are well suited, as different isotopologues tend to be well separated in frequency. Isotopic ratios of common elements like C, N, O, and S have been measured in various sources in the Milky Way (see, e.g., \citealp{wil94}) and some nearby galaxies (see, e.g., \citealp{omo07}). 

To probe chemical evolution on a cosmological timescale, we need observations at a range of redshifts. To this end, studies of molecules in absorption toward distant quasars have several advantages. There is no dilution by distance, hence this method can probe even rare isotopologues at high redshift. Unfortunately, there are only a handful of distant radio molecular absorbers currently known \citep{com08}, mostly limited by the number density of mm-bright quasars at high redshift ($z$$>$1), and the low probability of chance alignment between molecular clouds in a foreground galaxy and a background continuum source.

The only well-studied system to date is the $z$=0.89 molecular absorber (MA0.89) toward the lensed blazar PKS\,1830$-$211. A large number of molecules have been detected toward the SW line of sight, including a number of isotopologues which allowed for the measurement of isotopic ratios of C, N, O, S, Si, Cl, and Ar \citep{mul06,mul11,mul14,mul15}. Some of them were found to be quite different from the ratios obtained in the local interstellar medium (ISM), and consistent with the idea that at a redshift of $z$=0.89 (implying a maximum age of $\simeq 6$~Gyr) the interstellar gas has not been significantly polluted by the nucleosynthesis products of low mass stars. 

The only other similar molecular absorber known to date is the $z_{abs}$=0.68466 (heliocentric) absorber toward the lensed blazar B\,0218+357, discovered by \citet{wik95}. This absorber (hereafter MA0.68) has a maximum age of $\simeq 7$~Gyr, and appears as a nearly face-on spiral \citep{yor05}. The lensing produces two compact images (A and B) of the blazar separated by 0.3$''$, and an Einstein ring centered on image B, as seen at radio wavelengths \citep{pat93}. At mm wavelengths only the two compact images A and B are detected, owing to the steep spectral index of the Einstein ring. Molecular absorption is observed toward image A only, and species such as CO, CS, HCO$^+$, HCN, H$_2$O, NH$_3$, and H$_2$CO have been observed \citep{wik95,men96,com97,hen05,mul07,kan11}. The absorbing gas is found to have a density of a few 10$^2$~cm$^{-3}$ \citep{jet07} and a kinetic temperature $\sim$50~K \citep{hen05}, typical of Galactic diffuse clouds. According to the model of the lens by \cite{wuc04}, the absorption occurs at about 2~kpc from the intervening galaxy center.

This paper presents observations of HCO$^+$, HCN, CS, H$_2$S, and their isotopologues in MA0.68. This allows us, for the first time, to estimate the isotopic ratios $^{12}$C/$^{13}$C, $^{14}$N/$^{15}$N, $^{16}$O/$^{18}$O, $^{18}$O/$^{17}$O, and $^{32}$S/$^{34}$S in a second molecular absorber at intermediate redshift, and to compare them to the values found in MA0.89.

\section{Observations} 

Observations were done in the 3\,mm band with the IRAM Plateau de Bure Interferometer (PdBI) over different runs between 2005 and 2008 (see the journal of the observations in Table~\ref{tab:journal}). Each run lasted a few hours. The whole project was run as a back-up time filler when observing conditions were not good enough for regular imaging projects.

The bandpass response of the array was derived by observations of a bright radio quasar, such as 3C\,454.3 or 3C\,84. The 3\,mm spectral lines of HCO$^+$, HCN, CS, and H$_2$S, and their isotopologues were observed with 80~MHz-wide bands and $\sim$0.9~\kms\ velocity resolution. Continuum bands 320~MHz wide, with a coarse spectral resolution of 2.5~MHz, were used for calibration. The C$_2$H $N$=2--1 absorption was serendipitously detected in a continuum band.

All observations were taken in a compact configuration (except one observing run in November 2005 during the commissioning of a new very extended configuration of the array, see \citealt{mul08}), and the two lensed images of B\,0218+357, separated by 0.3$''$ were not spatially resolved. The absolute flux scale was derived from MWC\,349 whenever it was observed, or unset otherwise. The target visibilities were self-calibrated using a single point source model. Before 2007, the spectra were observed in single polarization. The installation of a new generation of receivers in 2007 allowed us to observe in dual polarization mode with an increase in observing efficiency and sensitivity. The resulting spectra were extracted directly from the averaged visibilities, and normalized to the total continuum flux density of the blazar (i.e., $I_A$+$I_B$=1).

\section{Analysis and results}

\subsection{Continuum illumination}

The two lensed images of B\,0218+357 cannot be spatially resolved in
our present data. We know, however, that the molecular absorption occurs only
in front of image A \citep{men96,mul07}. In order to convert the line
absorption depths to opacities, we need to know the flux ratio
\Rab=$I_A/I_B$ between the two compact lensed images of the blazar,
and the covering factor of source A, $f_c$, for the absorbing gas.

\begin{figure}[t]
\includegraphics[width=8.5cm]{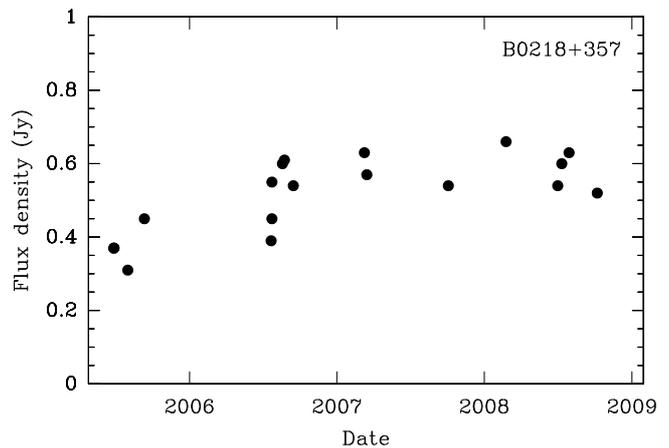}
\caption{Measurements of the total flux density of B\,0218+357 (A+B images) in the 3\,mm band. The absolute flux accuracy is expected to be of order 10--20\,\%.}
\label{fig:flux}
\end{figure}

The two images have been resolved in radio-cm observations, and their flux ratio changes with frequency \citep{mit06}. This is interpreted as the effect of free-free absorption by some ionised material in the line of sight \citep{mit07}. During the commissioning phase of a new very extended configuration of the PdBI (Nov. 2005), \citet{mul07} estimated \Rab=4.2$^{+1.8}_{-1.0}$ at 105~GHz, in agreement with the measurement at 15~GHz and model from \cite{mit07}. \cite{mar16} note, however, that during a VLA monitoring at 8~GHz and 15~GHz between 1996 October and 1997 January \citep{big99}, the flux ratio \Rab\ showed significant variations with time. These variations most likely result from intrinsic flux-density variations of the blazar modulated by the time delay between the two lensed images. In this model, even larger variations of the flux ratio can happen at higher frequencies (i.e., at mm wavelengths), such as those observed by \cite{mar13} toward PKS\,1830$-$211. Nevertheless, the total mm flux density variations of B\,0218+357 between 2005--2008 (Fig.\ref{fig:flux}) are smooth and on a timescale much longer than the time delay (10.5$\pm$0.4~days, \citealt{big99}), so that we do not expect that this affects the conclusions of this paper. 

\subsection{Stability of the absorption profile}

\begin{figure*}[ht]
\includegraphics[width=\textwidth]{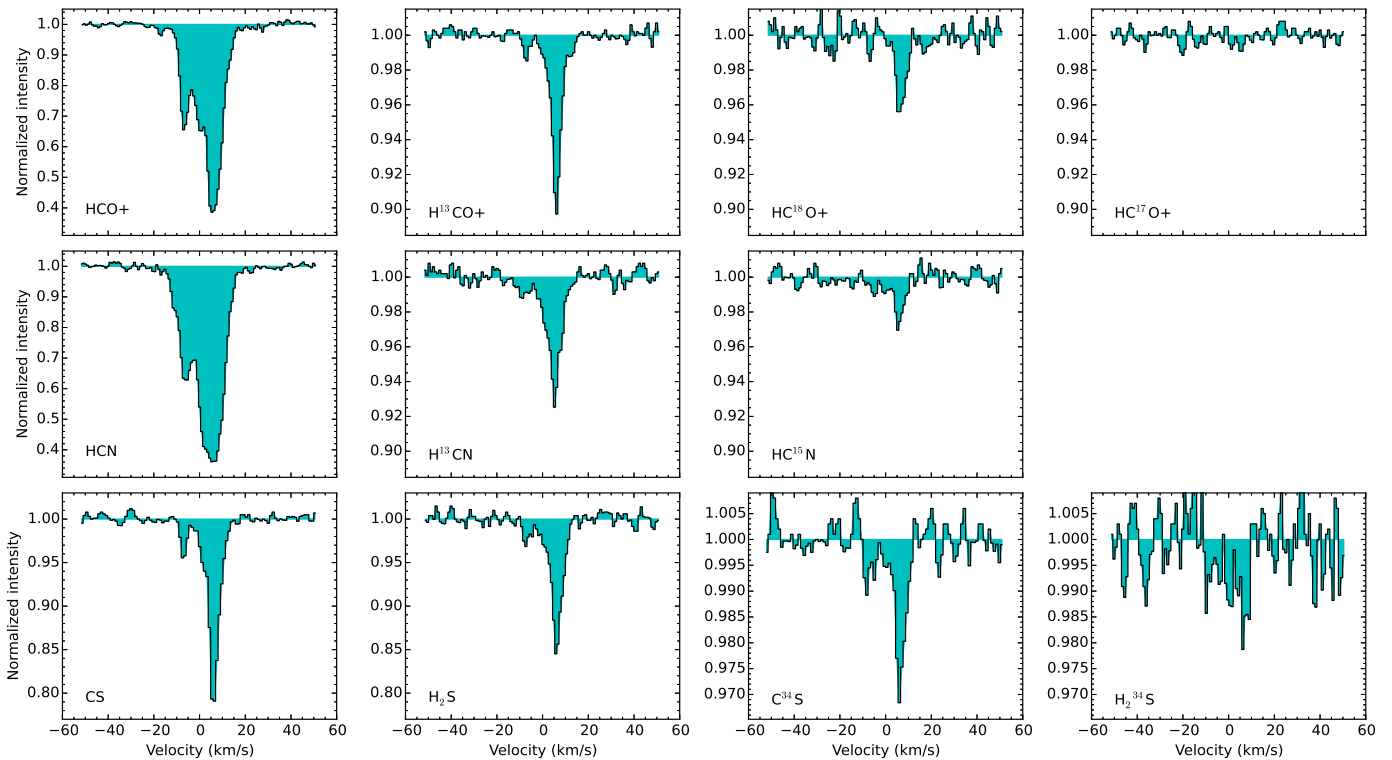}
\caption{Averaged spectra of the HCO$^+$, HCN, CS, and H$_2$S isotopologues. The velocity resolution is 0.9~\kms. The HCN spectrum is slightly broadened by the hyperfine structure.}
\label{fig:allspec}
\end{figure*}

The lensing and absorber system toward B\,0218+357 has similar properties to the one toward PKS\,1830$-$211. After the observations of drastic variations in the absorption line profile toward PKS\,1830$-$211 over a timescale on the order of a few months (see the
discussion in \citealt{mul08})\footnote{The variations in the
absorption profile toward PKS\,1830$-$211 are likely due to the
apparent highly superluminal motions ($v\sim8c$) of bright plasmons
injected in the blazar's jet, whose velocity is magnified by favorable
orientation of the jet close to the line of sight, by gravitational
lensing, and/or by secular precession of the jet.}, we decided to monitor
the absorption profile of the HCO$^+$ and HCN $J$=2--1 lines toward
B\,0218+357. The spectra, shown in Figs.\ref{fig:survey-spec-hco} and
\ref{fig:survey-spec-hcn}, do not show absorption variations down to a few
percent of the total continuum level. We therefore averaged all
the data of Table~\ref{tab:journal} for each species to obtain the
final spectra presented in Fig.~\ref{fig:allspec}.

\subsection{Absorption line profile} \label{sec:line-profile}

The absorption line profile is composed of two distinct features, the main one around a velocity of +6~\kms, and a second weaker peak at $-7$~\kms\ (Fig.~\ref{fig:allspec}; the velocities are given in the heliocentric frame, taking $z_{abs}$=0.68466 from \citealt{wik95}). The comparison of the H$^{12}$CO$^+$ and H$^{12}$CN $J$=2--1 line profiles with those of other molecules, including their rare H$^{13}$CO$^+$ and H$^{13}$CN isotopologues, suggests that they are saturated. Indeed, they reach roughly the same absorption depth of $\sim$0.6 below the total normalized continuum level, whereas the H$^{13}$CO$^+$ and H$^{13}$CN absorptions only reach depths around 0.1. The depth ratio of the $-7$~\kms\ to +6~\kms\ velocity components is also much lower for H$^{12}$CO$^+$ and H$^{12}$CN than for other species, further implying that the main component is saturated. The saturation of the HCN line gives $I_A \times f_c$=0.64 (the product $I_A \times f_c$ is degenerate in our data), so if we adopt \Rab=4.2, this implies $I_A$=0.81 and $f_c$=0.8. Furthermore, in the case of optically thin lines, the uncertainties in \Rab\ and $f_c$ do not significantly affect line opacity ratios. Hence, we convert the absorption spectra $I_v$ into opacity spectra:
\begin{equation}
 \tau_v = - \ln{ \left ( 1-\frac{I_v}{I_A \times f_c} \right ) }
\label{eq:tau}
\end{equation}
\noindent with $I_A \times f_c$=0.64. 

To derive a common normalized line profile we select all the lines that are optically thin with no hyperfine structure. We assume the same excitation, continuum illumination, fractional abundance, isotopic ratio, and source covering factor for all velocity components. The shape of the absorption profile suggests that a set of three Gaussian components (including a broad and a narrow feature for the dominant +6~\kms\ component) already provides a good solution. In Table~\ref{tab:gaussfit} we give the best fit solution of these three Gaussian components to the opacity spectra. The residuals of this fit are shown in Fig.~\ref{fig:spec-opa} and are within the noise.

\begin{table}[h]
\caption{Decomposition of our adopted line profile into Gaussian components. The areas are normalized to that of the first component.}
\label{tab:gaussfit}
\begin{center} \begin{tabular}{ccccc}
\hline
Gaussian & $v_0$ & FWHM & Area \\
         & (\kms) & (\kms) & (\kms) \\
\hline
\#1 & 6.176 (0.088) & 8.479 (0.346) & 1 \\
\#2 & 6.193 (0.039) & 2.907 (0.151) & 0.457 (0.063) \\ 
\#3 & $-$6.966 (0.118) & 3.261 (0.279) & 0.156 (0.013) \\
\hline
\end{tabular} \end{center}
\end{table}

\subsection{Isotopic ratios}

\begin{figure}[!htb]
    \centering
    \begin{minipage}{\columnwidth}
        \centering
        \includegraphics[width=\textwidth]{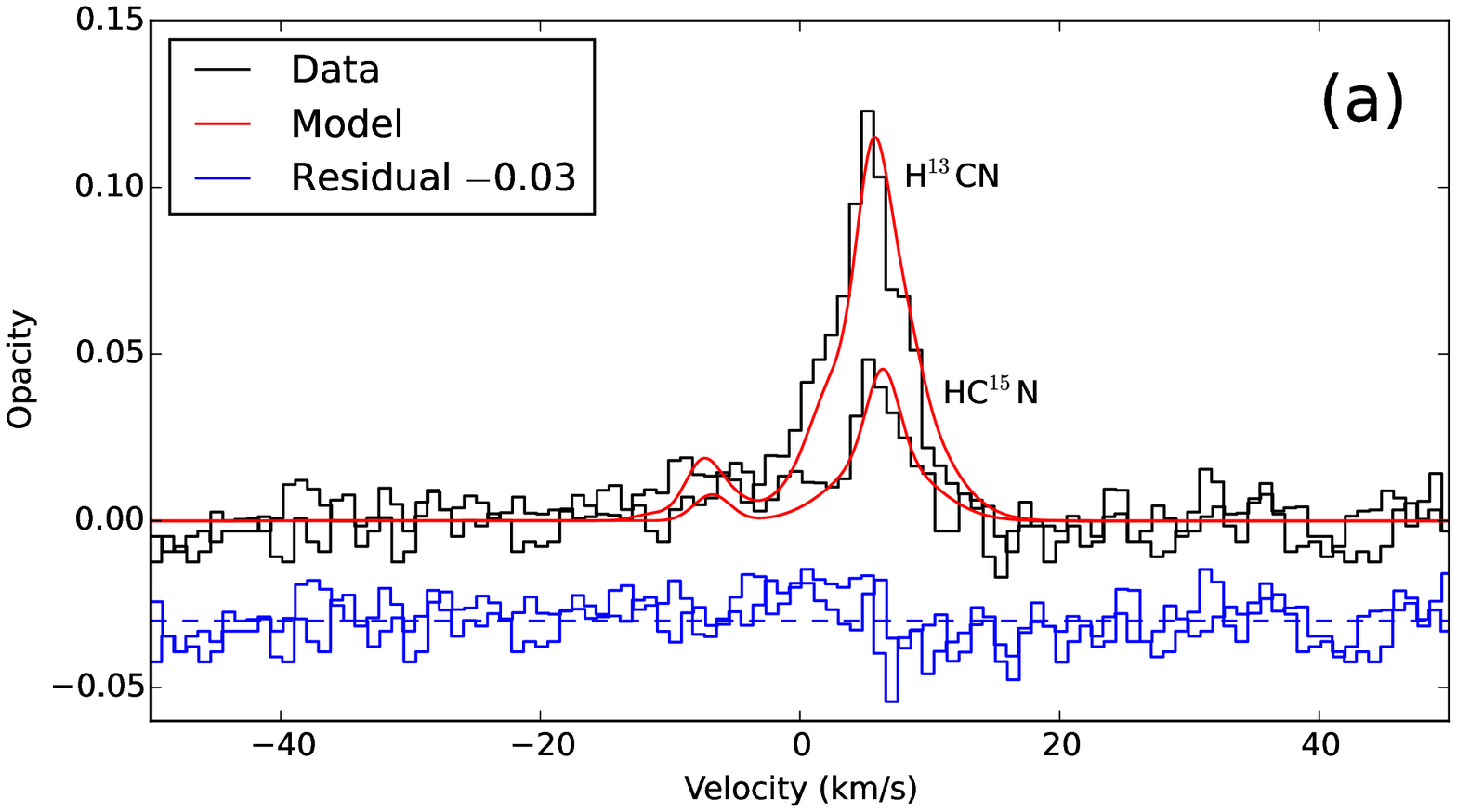} \\
        \includegraphics[width=\textwidth]{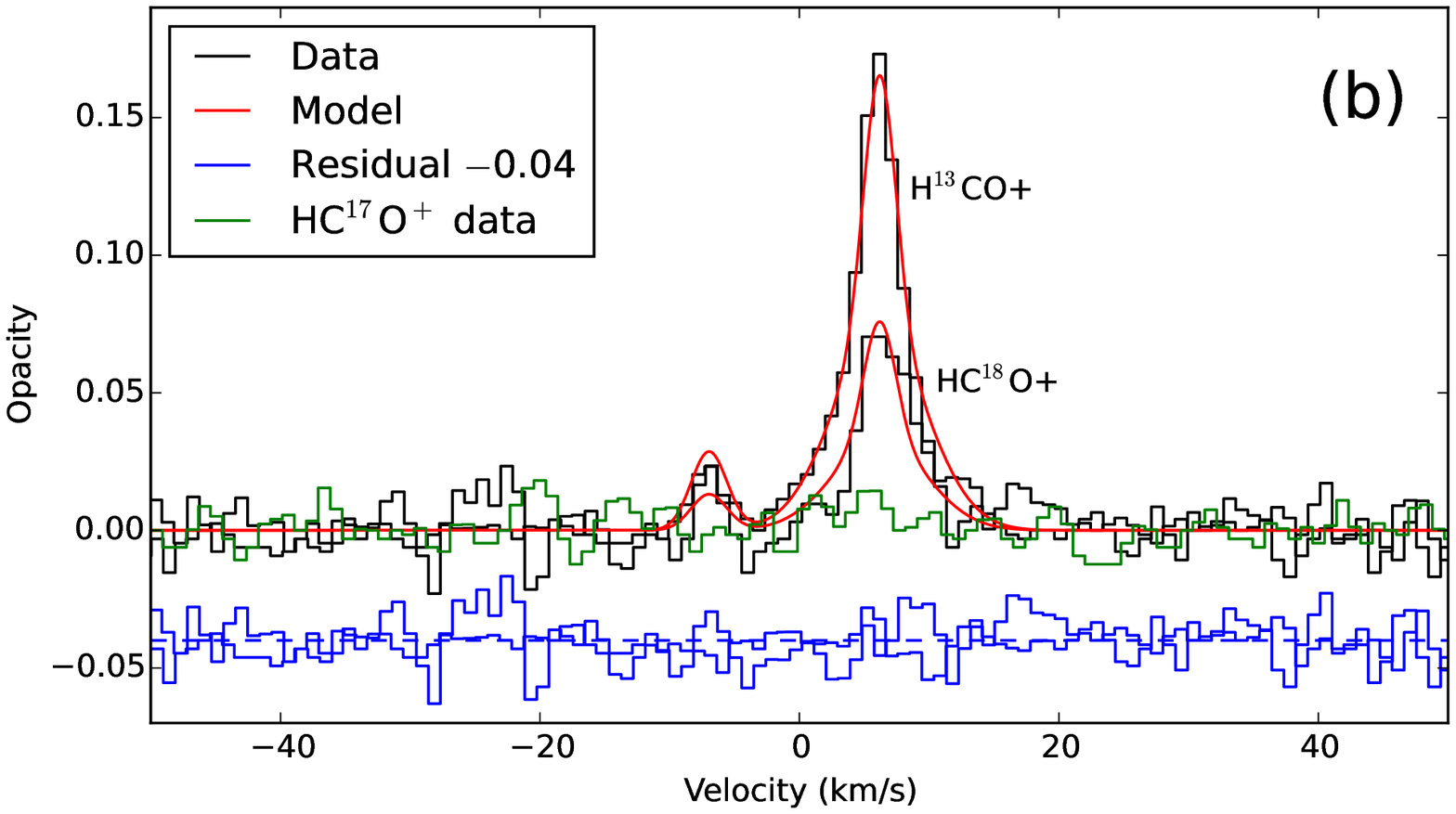} \\
        \includegraphics[width=\textwidth]{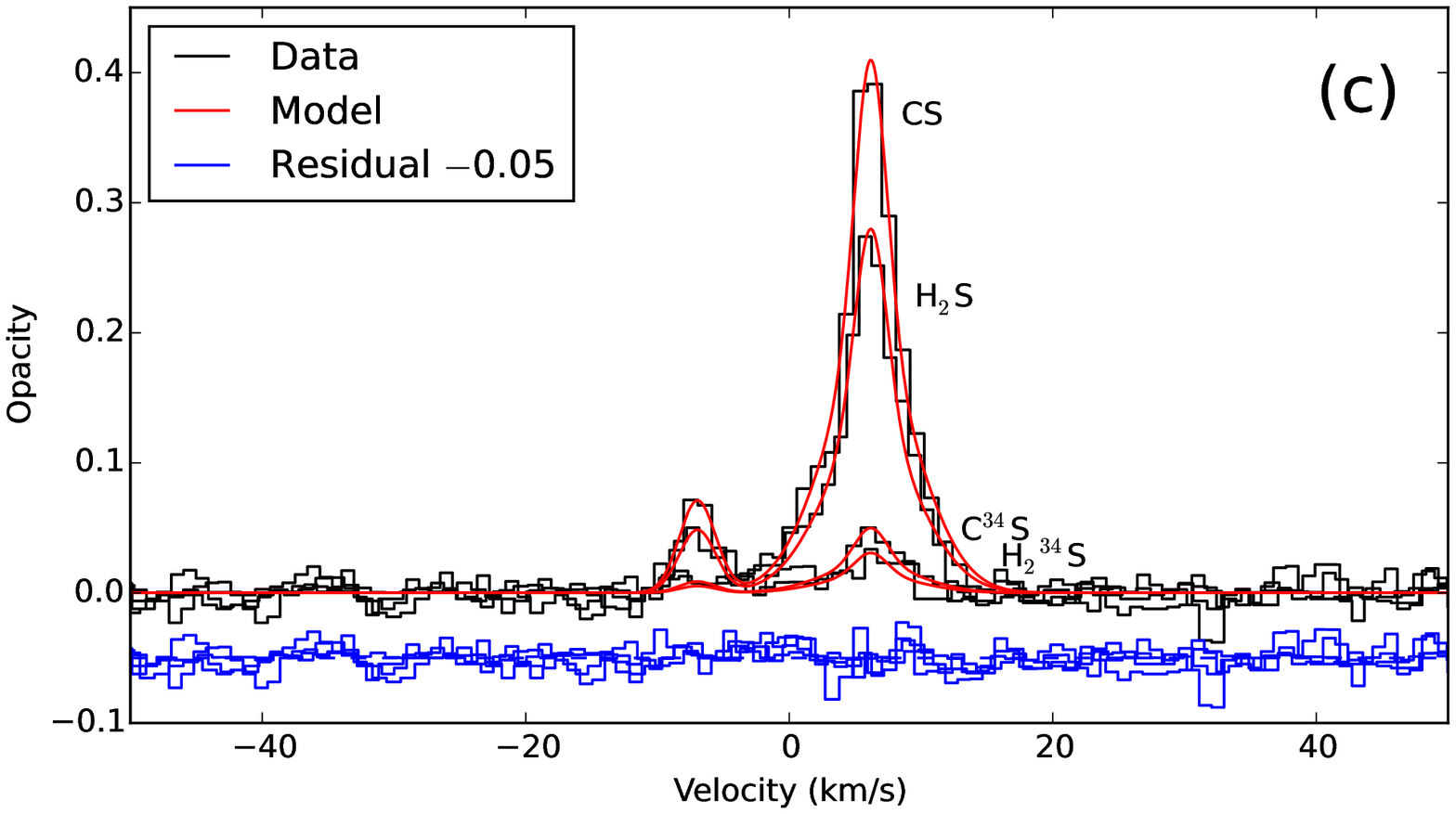}
    	\caption{Opacity spectra in black, model fits in red, and the residual (with some offset) in blue. In (a), hyperfine structure was taken into account in the fit. In (b), the spectrum of HC$^{17}$O$^+$, a non-detection, is shown in green.}
    	\label{fig:spec-opa}
    \end{minipage}
\end{figure}

The opacity spectra of the optically thin lines are shown in Fig.~\ref{fig:spec-opa}; the fits are shown in red. The normalized profile given in Table~\ref{tab:gaussfit} was fit to each line with a scaling factor. For the H$^{13}$CN and HC$^{15}$N isotopologues, the hyperfine structure is also taken into account, with the relative strengths expected when the sublevels are populated in proportion to their statistical weights. From the derived integrated opacities, we then calculate the column densities for the different species in our data. Since the absorbing gas is rather diffuse, we assume that collisional excitation is negligible compared to excitation by cosmic microwave background photons, hence that the excitation temperature is locked to the cosmic microwave background (CMB) temperature of 4.6~K at $z$=0.68 (see \citealp{mul13}). The calculated column densities are given in Table~\ref{tab:species} for species with optically thin lines.

The column densities (Table~\ref{tab:species}) and the resulting abundance ratios become very uncertain when the line opacities are large, which is the case for the main velocity components of HCO$^+$ and HCN main isotopomers, as discussed above. The difficulty may be removed by switching to the rare $^{13}$C, $^{15}$N, and $^{18}$O isotopologues, or by considering only the satellite velocity component at $-$7~\kms.

The $R_{\rm HCN}$=[H$^{13}$CN]/[HC$^{15}$N] and $R_{{\rm HCO}^+}$=[H$^{13}$CO$^+$]/[HC$^{18}$O$^+$] ratios can be determined with relatively good accuracy and should be free from opacity effects. Their values are $R_{\rm HCN}$=3.0$\pm$0.5 and $R_{{\rm HCO}^+}$=2.1$^{+0.4}_{-0.3}$, where uncertainties are given at 1$\sigma$ (68\% confidence level). While the former ratio $R_{\rm HCN}$ is similar to the double isotopic ratio $^{13}$C$^{14}$N/$^{12}$C$^{15}$N=3.0 in the solar system, $R_{{\rm HCO}^+}$ is definitely smaller in MA0.68 than in the solar system (5.5) or in the Milky Way ISM ($\gtrsim$10; see, e.g., Table~7 of \citealt{mul06}). Similarly, our non-detection of the HC$^{17}$O$^+$ line yields a firm lower limit HC$^{18}$O$^+$/HC$^{17}$O$^+$$>$7.5 (with a 3$\sigma$ confidence level, and neglecting the weakest HC$^{18}$O$^+$ component at $-$7 \kms). This limit is definitely larger than the $^{18}$O/$^{17}$O isotopic ratio in the solar system (5.4; \citealt{lod03}) and elsewhere in the Milky Way ISM ($2.88\pm0.11$ near the Galactic Center, $4.16\pm0.09$ across the Galactic disk out to a Galactocentric distance of $\sim 10$~kpc, and $5.03\pm0.46$ in the outer Galaxy at Galactocentric distances of 16--17~kpc; \citealt{wou08}). Finally, the $^{32}$S/$^{34}$S intensity ratios for the CS and H$_2$S isotopologues (combined value 8.1$_{-1.1}^{+1.4}$) are both less than half the values of the $^{32}$S/$^{34}$S isotopic ratio observed in the solar system (22; \citealt{lod03}) or the general Milky Way ISM ($\simeq$20; \citealt{chi96,luc98}).

The molecular isotopic ratios may be affected by chemical fractionation (see \citealt{rou15} and references therein). This is particularly the case for C-bearing molecules and, to a lesser degree, N-bearing molecules in cold clouds, where the abundance of the heavier $^{13}$C and $^{15}$N isotopologues may be enhanced relative to their main $^{12}$C and $^{14}$N counterparts. We note, however, that isotopic fractionation, if present, would act to decrease the measured $^{12}$C/$^{13}$C isotopic ratio and hence increase the double ratio ($^{16}$O/$^{18}$O)/($^{12}$C/$^{13}$C), relative to their true values. The true value of the double ratio would then be even smaller than our measurement of $R_{{\rm HCO}^+}$=2.1, and hence more different from the solar system and Milky Way ratios.

As concerns the $^{12}$C/$^{13}$C isotopic ratio, the opacity ratio of the weak v=$-$7~\kms\ component yields [H$^{12}$CO$^+$]/[H$^{13}$CO$^+$]=38$\pm$5, clearly smaller than the solar system value of 89 and close to $^{12}$C/$^{13}$C in MA0.89, but may still be underestimated owing to fractionation.

\begin{table}[h]
\caption{Isotopic ratios in MA0.68, and MA0.89.}
\label{tab:isotopic-ratios}
\begin{center} \begin{tabular}{ccc}
\hline
Ratio & MA0.68 & MA0.89 \\ 
\hline 
H$^{13}$CN / HC$^{15}$N & $3.0\pm 0.5$ & $4.8\pm 0.2$$^\ddagger$  \\ 
H$^{13}$CO$^+$ / HC$^{18}$O$^+$ & $2.1^{+0.4}_{-0.3}$ & $2.15 \pm 0.04$$^\ddagger$  \\ 
HC$^{18}$O$^+$ / HC$^{17}$O$^+$ & $>7.5$ & $13\pm 3$$^\ddagger$  \\
CS / $^{34}$CS & $7.9^{+1.6}_{-1.4}$ & $10.4^{+0.8}_{-0.7}$$^\dagger$ \\   
H$_2$S / H$_2$$^{34}$S & $8.6^{+2.4}_{-1.7}$ & $8\pm 1.5$$^\dagger$ \\  

\hline
\end{tabular} \end{center}
\mbox{\,} \vskip -.4cm
\hspace{0.9cm} $^\dagger$\citet{mul06}

\hspace{0.9cm} $^\ddagger$Averaged result from \citet{mul11}
\end{table}

\begin{figure}[h]
\includegraphics[width=8.5cm]{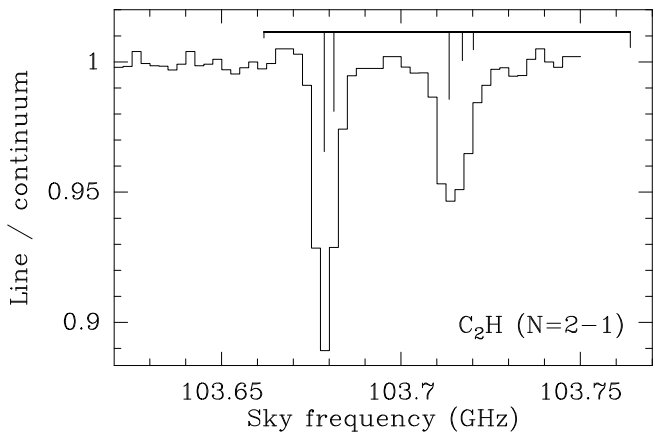}
\caption{Spectrum of the C$_2$H $N$=2--1 absorption, with a spectral resolution of 2.5~MHz.
The spectrum follows the relative strengths expected at local thermodynamic equilibrium for optically thin hyperfine structure components.
}
\label{fig:c2h}
\end{figure}

We can also use the low resolution C$_2$H $N$=2--1 absorption profile (Fig.~\ref{fig:c2h}) to estimate the H$_2$ column density, assuming that the relative [C$_2$H]/[H$_2$] and [HCO$^+$]/[H$_2$] abundances are comparable to those in Galactic diffuse clouds and in the MA0.89 NE absorption component toward PKS\,1830$-$211 \citep{luc00, mul11}. First of all, we note that the C$_2$H $N$=2--1 line absorption is weak, hence presumably optically thin, and that the relative intensities of the two fine-structure components follow the local thermodynamic equilibrium value. Taking [C$_2$H]/[H$_2$]=3$\times$10$^{-8}$, we obtain a total H$_2$ column density of $\sim$10$^{22}$~cm$^{-2}$. In turn, we can roughly estimate a HCO$^+$ column density of $\sim$3$\times$10$^{13}$~cm$^{-2}$, using [HCO$^+$]/[H$_2$]=6$\times$10$^{-9}$. Comparing this value with that measured for H$^{13}$CO$^+$ yields a ratio [HCO$^+$]/[H$^{13}$CO$^+$]$\sim$30, a value which is uncertain but similar to that derived above for the v=$-$7~\kms\ component and to the $^{12}$C/$^{13}$ C ratio derived for MA0.89 SW from the fit of two different rotational transitions of HCO$^+$, HCN, and HNC by \citealt{mul11}.

Adopting $^{12}$C/$^{13}$C$\sim$40, we obtain $^{14}$N/$^{15}$N$\sim$120 and $^{16}$O/$^{18}$O$\sim$80 from the double isotopologue ratios in Table~\ref{tab:isotopic-ratios}.

\section{Discussion}

\begin{figure}[t]
\begin{minipage}{\linewidth}
\includegraphics[width=\linewidth]{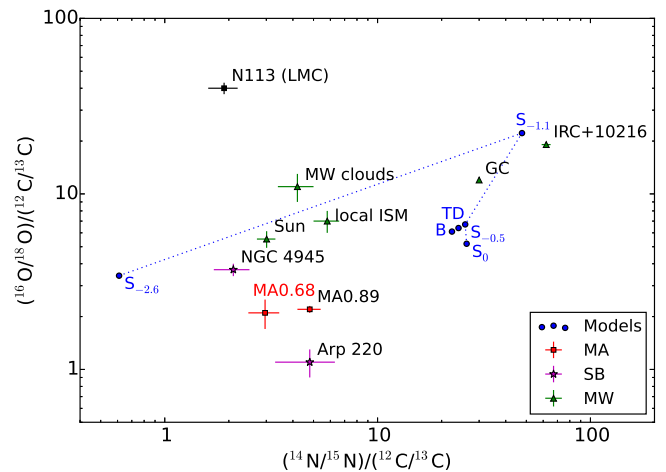}
\caption[]{$^{16}$O/$^{18}$O vs. $^{14}$N/$^{15}$N ratios, both normalized by the $^{12}$C/$^{13}$C ratio, for various astronomical sources. Our result for MA0.68 is in red, as is MA0.89 (MA), starburst galaxies are magenta stars (SB), Galactic sources are green triangles (MW), and the predictions from chemical evolution models in the solar neighborhood by \cite{kob11} are in blue and labeled as S$_{X}$, where $X$ denotes the metallicity [Fe/H]. Their models for the bulge (B) and thick disk (TD), also in blue, are for [Fe/H]=-0.5. 
\footnotetext{MA0.68: This work. MA0.89: \citet{mul11}. NGC4945: \citet{wan04}. Arp 220: \citet{wan16,mar11}. N113: \citet{chi99b,wan09}. IRC+10216: \citet{he08}. MW clouds: \citet{luc98}. GC, local ISM: \citet{wil94}. Sun: \citet{and89}.}}
\label{fig:2d}
\end{minipage}
\end{figure}

\begin{figure}[t]
\begin{minipage}{\linewidth}
\includegraphics[width=\linewidth]{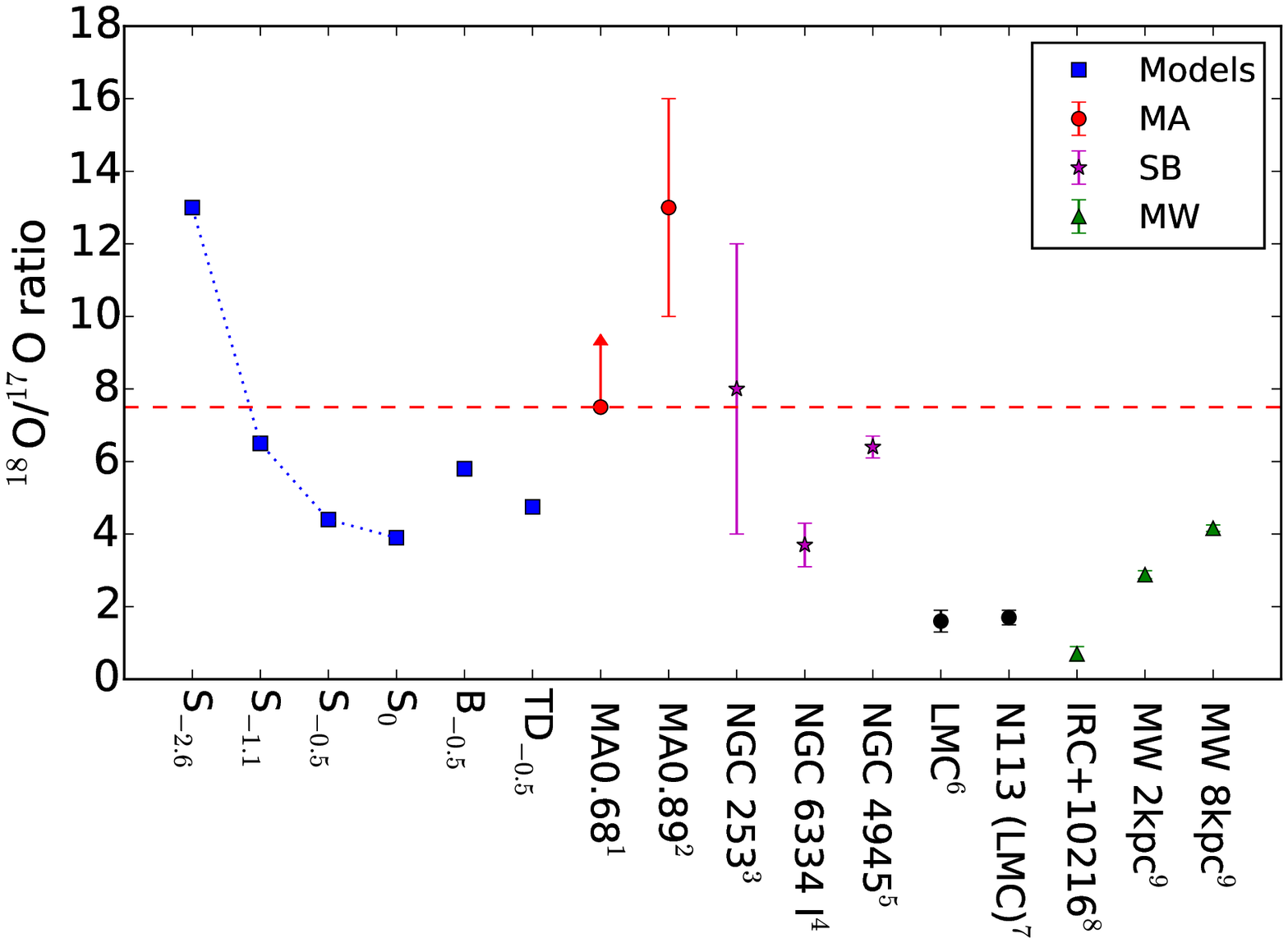}
\caption[]{$^{18}$O/$^{17}$O ratios in various sources and models, colored and labeled as in Fig.~\ref{fig:2d}.
\footnotetext{$^1$This work. $^2$\citet{mul11}. $^3$\citet{hen14}. $^4$\citet{emp10}. $^5$\citet{wan04}. $^6$\citet{hei99}. $^7$\citet{wan09}. $^8$\citet{kah92}. $^9$\citet{wou08}.}}
\label{fig:1817}
\end{minipage}
\end{figure}

\begin{figure}[t]
\begin{minipage}{\linewidth}
\includegraphics[width=\linewidth]{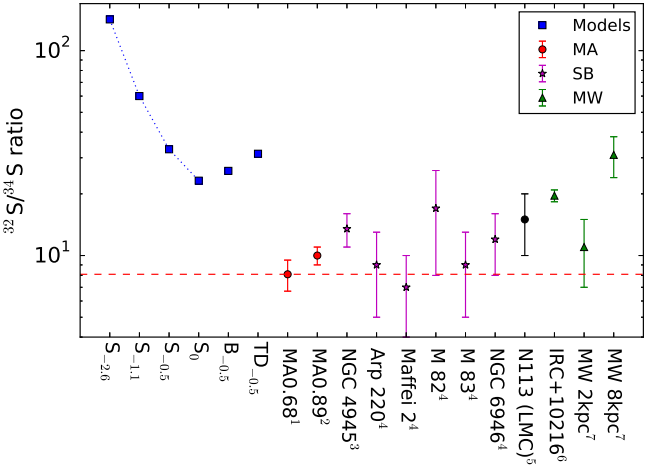}
\caption[]{$^{32}$S/$^{34}$S ratios in various sources and models, colored and labeled as in Fig.~\ref{fig:2d}.
\footnotetext{$^1$This work. $^2$\citet{mul06}. $^3$\citet{wan04}. $^4$\citet{mar09}. $^5$\citet{wan09}. $^6$\citet{pat09}. $^7$\citet{chi96}.}}
\label{fig:3234}
\end{minipage}
\end{figure}

We compare our measurements in MA0.68 to the isotopic ratios observed in other sources, including MA0.89 (the molecular absorber at z=0.89 toward PKS\,1830$-$211), nearby starburst galaxies, and Galactic sources. We also compare them to model predictions by \citet{kob11} for the solar neighborhood, Galactic bulge, and thick disk at different metallicities. These comparisons are shown in Figs.~\ref{fig:2d}, \ref{fig:1817}, and \ref{fig:3234}.

First, we emphasise that all measured isotopic ratios in MA0.68 are similar to those found in MA0.89. This is interesting because the two absorbers are not connected, but have approximately the same look-back time of 6-7~Gyr, and absorption occurs at a galactocentric radius of $\sim$2~kpc in both. However, we do not have information about the metallicities and star formation histories of these two absorbers. 

We find that the isotopic ratios in MA0.68 and MA0.89 deviate significantly from the values measured in the solar neighborhood. However, for $^{12}$C/$^{13}$C in MA0.68 our estimated value of $\sim$40 is consistent with the values reported by \citet{sav02} and \citet{mil05} at a galactocentric distance (D$_\mathrm{GC}$) of 2~kpc in the Milky Way. Between 2~kpc and 8~kpc, the $^{12}$C/$^{13}$C ratio is found to increase by a factor of $\sim$2 \citep{sav02,mil05}. Similarly for $^{32}$S/$^{34}$S, our value of 8.1 is consistent with the values at D$_\mathrm{GC}$=2~kpc derived by \citet{chi96}, assuming that the above mentioned galactocentric gradient in $^{12}$C/$^{13}$C is valid. On the other hand, taking $^{12}$C/$^{13}$C$\sim$40 in MA0.68, we find ratios of $\sim$120 and 80 for $^{14}$N/$^{15}$N and $^{16}$O/$^{18}$O, respectively, which are lower than the corresponding values in the Milky Way at D$_\mathrm{GC}$=2 kpc by a factor of 2--3 \citep{ada12,wil94}.

For $^{18}$O/$^{17}$O, our lower limit in MA0.68 and measurement in MA0.89 are much larger than the remarkably constant values observed in the Milky Way, including at D$_\mathrm{GC}$=2~kpc \citep{wou08}. Neither opacity, nor fractionation are expected to affect the measurements of $^{18}$O/$^{17}$O and the values should genuinely result from nucleosynthetic history. Both oxygen ratios are indicative of enrichment on short timescales as $^{18}$O is mainly produced by He-burning in massive stars. 

The closest match for the ratios observed in MA0.68 and MA0.89 are nearby starburst galaxies. We expect that both are dominated by the products of massive stars: MA0.68 and MA0.89 because of their youth, and starburst galaxies because they are actively forming stars, the most massive of which will enrich the ISM the fastest. There are, however, large observational uncertainties and apparent source-to-source variations in the mostly weak and very broad line profiles of starburst galaxies. Furthermore, extragalactic observations are often averaged over a larger area than is sampled by the line of sight through MA0.68, making detailed comparisons difficult.

It is interesting to compare our results in MA0.68 to chemical evolution models, although there are very few which report the abundances of different isotopes over time. Hence, we consider the models by \citet{kob11}, which attempt to reproduce the chemical evolution and observed metallicity distribution in different parts of the Milky Way: the solar neighborhood, bulge, and thick disk. The models include enrichment from AGB stars, massive evolved stars, core collapse supernovae (Type II SNe), and single degenerate Type Ia SNe. They give isotopic abundances at a few different metallicities, which correspond to different epochs of enrichment. At [Fe/H] = $-$2.6, enrichment is mainly from Type II SNe. At [Fe/H] = $-$1.1, AGB stars start to enrich the ISM, along with Type II SNe. At [Fe/H] = $-$0.5, there is enrichment from Type II SNe, AGB stars, and Type Ia SNe. And [Fe/H] = 0 is the present solar neighborhood. In Figs.~\ref{fig:2d}, \ref{fig:1817}, and \ref{fig:3234} the points from the \citet{kob11} solar neighborhood models are labeled as S$_{X}$, where $X$ denotes the metallicity [Fe/H]. The bulge model is labeled B$_{X}$, and the thick disk model TD$_{X}$.

Figure~\ref{fig:2d} shows the predicted $^{16}$O/$^{18}$O and $^{14}$N/$^{15}$N ratios, both normalized by $^{12}$C/$^{13}$C. In $^{14}$N/$^{15}$N, MA0.68 falls between the models S$_{-2.6}$ and S$_{-1.1}$. However, in $^{16}$O/$^{18}$O we see values lower than all model predictions. In the $^{18}$O/$^{17}$O ratio we find a lower limit of 7.5 in MA0.68, consistent only with the model S$_{-2.6}$. This agrees with enrichment mainly from Type II SNe as $^{18}$O is mainly produced by He-burning in massive stars, while large amounts of $^{17}$O are produced in AGB stars.  

For the $^{32}$S/$^{34}$S ratio our value for MA0.68 is much lower than the models of \citet{kob11} predict for all metallicities. A possible source of the low $^{32}$S/$^{34}$S ratio is hypernovae, core collapse supernovae with explosion energies more than 10 times that of regular supernovae (i.e., $>$10$^{52}$ ergs), which produce a lot of $^{34}$S. The fraction of hypernovae is not well constrained, so \citet{kob11} assume a fraction of 0.5 for M $\geqslant$ 20 \msun. Hence the observed low $^{32}$S/$^{34}$S ratios might imply that hypernovae are more common than assumed, at least in certain environments such as MA0.68 and starburst galaxies.

\section{Summary and conclusions} 

We present measurements of isotopic ratios of C, N, O, and S elements in the $z$=0.68 molecular absorber toward B\,0218+357. To this end, we use several isotopologues of HCO$^+$, HCN, CS, and H$_2$S. The analysis of the line profiles helps us to constrain the continuum illumination and source covering factor.

The $^{12}$C/$^{13}$C, $^{14}$N/$^{15}$N, and $^{16}$O/$^{18}$O ratios are difficult to measure separately owing to opacity effects. However, we obtain the robust and accurate double isotopic abundance ratios [H$^{13}$CN]/[HC$^{15}$N]=$3.0\pm 0.5$ and [H$^{13}$CO$^+$]/[HC$^{18}$O$^+$]= $2.1^{+0.4}_{-0.3}$. Discarding possible fractionation effects, these abundance ratios can be converted into the $^{14}$N/$^{15}$N and $^{16}$O/$^{18}$O isotopic ratios normalized to $^{12}$C/$^{13}$C, respectively.  Using chemical abundance arguments, we argue that $^{12}$C/$^{13}$C should be close to a value of $\sim$40, implying $^{14}$N/$^{15}$N$\sim$120 and $^{16}$O/$^{18}$O$\sim$80, although these ratios might be uncertain by a factor of a few. On the other hand, we measure a remarkably low $^{32}$S/$^{34}$S ratio of 8.1$_{-1.1}^{+1.4}$ from optically thin lines of CS and H$_2$S, and obtain a large lower limit of 7.5 (3$\sigma$) for $^{18}$O/$^{17}$O, from our non-detection of HC$^{17}$O$^+$. The last two ratios are expected to be free of opacity and fractionation effects.

All measured isotopic ratios in MA0.68 are similar to those found in the $z$=0.89 molecular absorber toward PKS\,1830$-$211, and they all differ from values in the solar neighborhood. In $^{12}$C/$^{13}$C and $^{32}$S/$^{34}$S, the ratios found in MA0.68 appear similar to those in the Milky Way at the same galactocentric distance of 2~kpc. Both the $^{14}$N/$^{15}$N and $^{16}$O/$^{18}$O ratios seem to be significantly lower in MA0.68, and the $^{18}$O/$^{17}$O (3$\sigma$) lower limit is significantly higher than ratios measured in the Milky Way. This is indicative of enrichment mainly by massive stars, which produce proportionately more $^{18}$O and $^{15}$N.

This work shows that redshifted molecular absorbers are interesting targets to determine the evolution of isotopic ratios, and hence the enrichment history of the interstellar medium. However, the interpretation of the derived isotopic ratios is hampered by the poor knowledge we have of, for example, the location of the absorbing gas in the absorber, the gas metallicity, and the star formation history.

The discovery of new molecular absorbers such as PKS\,1830$-$211 and B\,0218+357, hopefully at $z$$>$1, would further constrain our knowledge of nucleosynthesis and chemical evolution, as well as our understanding of the origin and evolution of the elements in the Universe.

\begin{acknowledgements}
The authors thank the Plateau de Bure Observatory staff and IRAM-Grenoble SOG for their support in the observations. This work is based on observations carried out with the IRAM Plateau de Bure Interferometer. IRAM is supported by INSU/CNRS (France), MPG (Germany), and IGN (Spain).
We thank the referee for the extensive and useful comments which helped us to improve the clarity of the manuscript.
\end{acknowledgements}

\begin{appendix}

\section{Appendix}

\begin{figure}[h!]
\includegraphics[width=8.5cm]{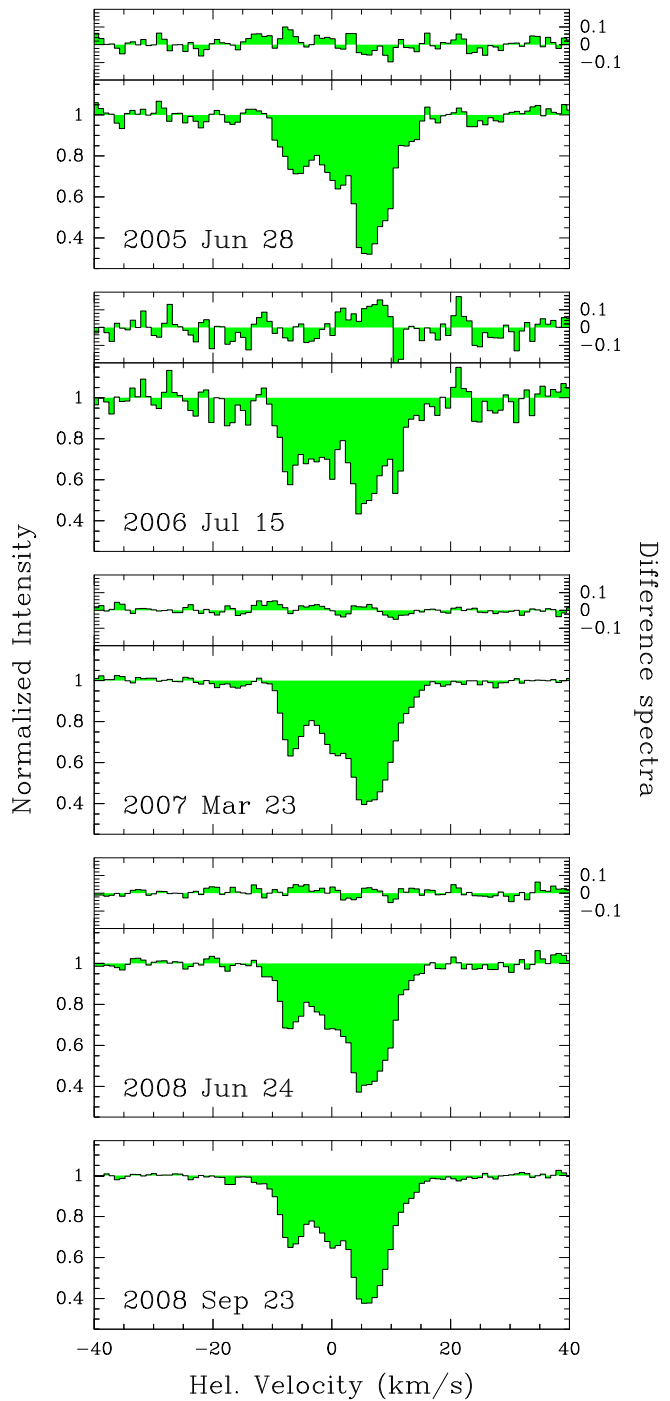}
\caption{Spectra of the HCO$^+$ $J$=2--1 absorption toward B\,0218+357 at different epochs between 2005 and 2008. Difference spectra are also shown with respect to the last (and best signal-to-noise ratio) spectrum obtained on 2008 Sept 23.}
\label{fig:survey-spec-hco}
\end{figure}

\begin{figure}[h!]
\includegraphics[width=8.5cm]{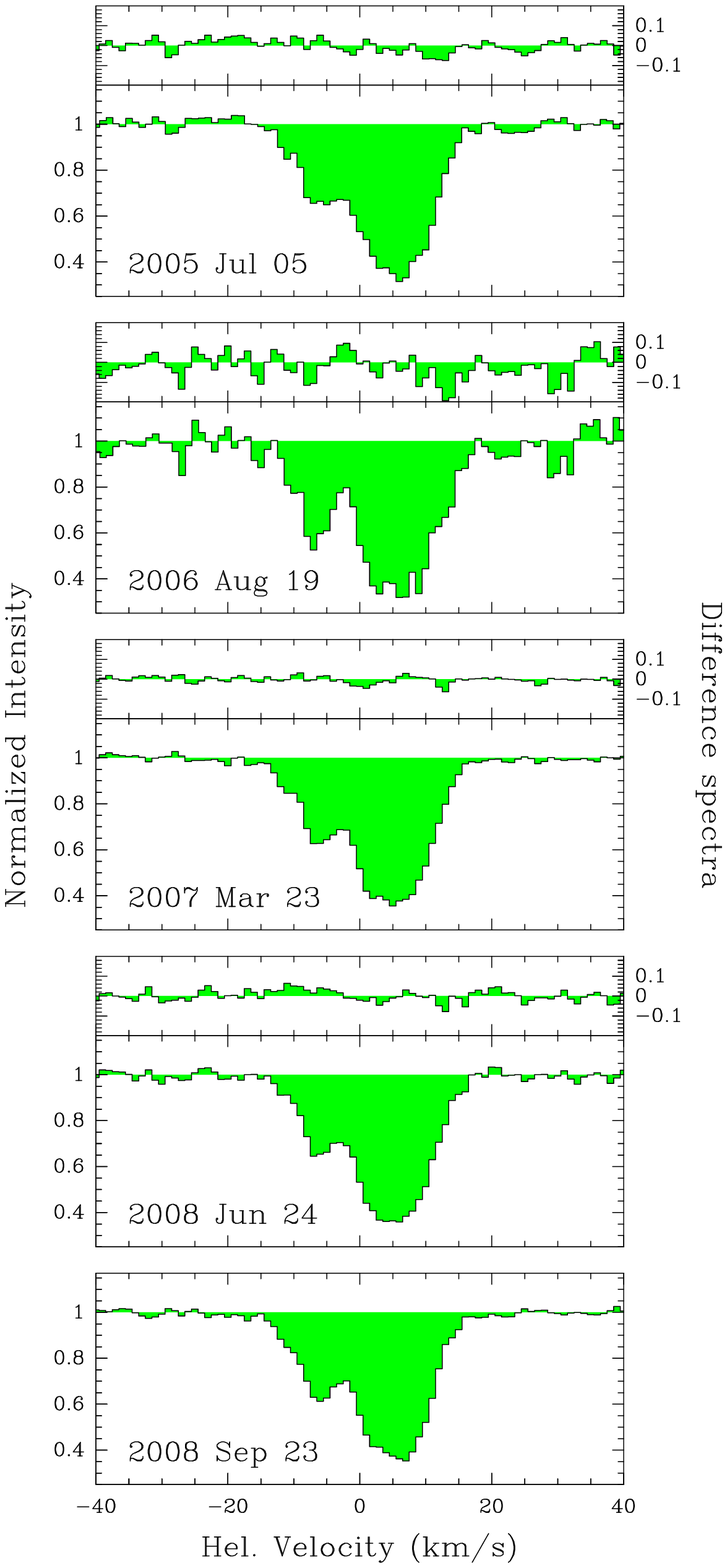}
\caption{Spectra of the HCN $J$=2--1 absorption toward B\,0218+357 at different epochs between 2005 and 2008. Difference spectra are also shown with respect to the last (and best signal-to-noise ratio) spectrum obtained on 2008 Sept 23.}
\label{fig:survey-spec-hcn}
\end{figure}

\begin{table*}[h]
\caption{Journal of PdBI observations toward \B0218.}
\label{tab:journal}
\begin{center} \begin{tabular}{cccc}
\hline
Date & Observed species & Note & Total flux density \\
     &                  &      &  A+B (Jy)   \\
\hline
 2005 Jun. 28 & HCO$^+$                        & & 0.37 \\
 2005 Jul. 04 & HCN                            & &  \\
 2005 Aug. 01 & H$^{13}$CO$^+$, H$^{13}$CN                 & &  0.31 \\
 2005 Aug. 18 & H$^{13}$CO$^+$, H$^{13}$CN                 & &   \\ 
 2005 Aug. 28 & H$^{13}$CO$^+$, H$^{13}$CN                 & &   \\ 
 2005 Sep. 03 & H$^{13}$CO$^+$, H$^{13}$CN                 & &   \\ 
 2005 Sep. 06 & H$^{13}$CO$^+$, H$^{13}$CN                 & &  \\
 2005 Sep. 11 & H$^{13}$CO$^+$, H$^{13}$CN                 & &  0.45 \\
 2005 Nov. 08 & HCO$^+$                      & Ext. configuration &  \\
 2006 Jul. 15 & HCO$^+$ & &  \\
 2006 Jul. 16 & H$^{13}$CO$^+$, HC$^{17}$O$^+$ & & \\ 
 2006 Jul. 19 & H$^{13}$CO$^+$, HC$^{17}$O$^+$ & & \\
 2006 Jul. 22 & H$^{13}$CO$^+$, HC$^{17}$O$^+$ & & 0.39 \\
 2006 Jul. 24 & HC$^{18}$O+                               & & 0.44 \\
 2006 Aug. 02 & H$^{13}$CO$^+$, HC$^{17}$O$^+$ & & 0.53 \\ 
 2006 Aug. 05 & H$^{13}$CO$^+$, HC$^{17}$O$^+$ & &  \\
 2006 Aug. 09 & H$^{13}$CO$^+$, HC$^{17}$O$^+$ & &  \\
 2006 Aug. 19 & HCN                            & & 0.60 \\
 2006 Aug. 24 & H$^{13}$CN, HC$^{15}$N                     & & 0.61 \\
 2006 Aug. 25 & HC$^{18}$O$^+$                               & &  \\
 2006 Sep. 09 & HC$^{18}$O$^+$                         & &  \\
 2006 Sep. 15 & HC$^{18}$O$^+$                               & & 0.54 \\
 2007 Feb. 20 & C$^{34}$S & Dual pol. &   \\ 
 2007 Feb. 28 & C$^{34}$S & Dual pol. &  \\
 2007 Mar. 01 & C$^{34}$S & Dual pol. &  \\
 2007 Mar. 08 & C$^{34}$S & Dual pol. &  \\
 2007 Mar. 10 & C$^{34}$S & Dual pol. & 0.63 \\
 2007 Mar. 15 & C$^{34}$S & Dual pol. &  \\
 2007 Mar. 16 & C$^{34}$S & Dual pol. & 0.57 \\
 2007 Mar. 23 & HCO$^+$, HCN & Dual pol. &  \\
 2007 Mar. 24 & C$^{34}$S & Dual pol. & \\
 2007 Jul. 01 & H$^{13}$CO$^+$, HC$^{18}$O$^+$, HC$^{17}$O$^+$ & & \\ 
 2007 Oct. 04 & H$^{13}$CO$^+$, HC$^{18}$O$^+$, HC$^{17}$O$^+$ & & 0.54 \\
 2007 Oct. 05 & H$^{13}$CO$^+$, HC$^{18}$O$^+$, HC$^{17}$O$^+$ & &  \\
 2007 Oct. 20 & H$^{13}$CO$^+$, H$^{13}$CN, HC$^{15}$N & &  \\ 
 2007 Dec. 06 & H$^{13}$CO$^+$, H$^{13}$CN, HC$^{15}$N, H$_2$S, H$_2$$^{34}$S & &  \\
 2008 Jan. 09 & H$^{13}$CO$^+$, H$^{13}$CN, HC$^{15}$N, H$_2$S, H$_2$$^{34}$S & &  \\ 
 2008 Feb. 21 & CS & Dual pol. &  \\
 2008 Feb. 24 & C$^{34}$S & Dual pol. & 0.66 \\
 2008 Feb. 25 & C$^{34}$S & Dual pol. &  \\
 2008 Feb. 27 & C$^{34}$S & Dual pol. &  \\
 2008 Mar. 04 & H$^{13}$CO$^+$, HC$^{17}$O$^+$, C$_2$H & Dual pol. &  \\
 2008 Mar. 15 & H$^{13}$CO$^+$, HC$^{17}$O$^+$, C$_2$H & Dual pol. &  \\
 2008 Jun. 24 & HCO$^+$, HCN & &  \\
 2008 Jun. 26 & H$^{13}$CO$^+$, H$^{13}$CN, HC$^{15}$N, H$_2$S, H$_2$$^{34}$S & & \\
 2008 Jun. 27 & H$^{13}$CO$^+$, H$^{13}$CN, HC$^{15}$N, H$_2$S, H$_2$$^{34}$S & &  \\
 2008 Jul. 01 & H$^{13}$CO$^+$, H$^{13}$CN, HC$^{15}$N, H$_2$S, H$_2$$^{34}$S & & 0.54 \\
 2008 Jul. 11 & H$^{13}$CO$^+$, H$^{13}$CN, HC$^{15}$N, H$_2$S, H$_2$$^{34}$S & & 0.60 \\
 2008 Jul. 12 & H$^{13}$CO$^+$, H$^{13}$CN, HC$^{15}$N, H$_2$S, H$_2$$^{34}$S & &  \\
 2008 Jul. 29 & H$^{13}$CO$^+$, H$^{13}$CN, HC$^{15}$N & Dual pol. & 0.63 \\
 2008 Aug. 04 & H$^{13}$CO$^+$, H$^{13}$CN, HC$^{15}$N & Dual pol. & \\
 2008 Aug. 08 & H$^{13}$CO$^+$, HC$^{17}$O$^+$, C$_2$H & Dual pol. &  \\
 2008 Aug. 17 & H$^{13}$CO$^+$, HC$^{17}$O$^+$, C$_2$H & Dual pol. &  \\
 2008 Sep. 11 & H$^{13}$CO$^+$, HC$^{17}$O$^+$, C$_2$H & Dual pol. &   \\
 2008 Sep. 12 & H$^{13}$CO$^+$, HC$^{17}$O$^+$, C$_2$H & Dual pol. &   \\
 2008 Sep. 23 & HCO$^+$, HCN & Dual pol. &   \\
 2008 Oct. 07 & H$^{13}$CO$^+$, HC$^{17}$O$^+$, C$_2$H & Dual pol. &  0.52 \\
 2008 Oct. 08 & H$^{13}$CO$^+$, HC$^{17}$O$^+$, C$_2$H & Dual pol. &  \\
 2008 Oct. 18 & H$^{13}$CO$^+$, HC$^{17}$O$^+$, C$_2$H & Dual pol. &  \\
 2008 Oct. 20 & H$^{13}$CO$^+$, H$^{13}$CN, HC$^{15}$N, H$_2$S H$_2$$^{34}$S & &  \\
 2008 Nov. 28 & H$^{13}$CO$^+$, H$^{13}$CN, HC$^{15}$N, H$_2$S H$_2$$^{34}$S & & 0.61 \\
\hline
\end{tabular} \end{center} \end{table*}

\begin{sidewaystable*}[htbp]
\caption{Observed transitions.}
\label{tab:species}
\begin{center} 
\begin{tabular}{ccccccc}
\hline
Line & Rest freq. & Redshifted       & Integrated  & Channel rms noise $^c$ & $\alpha$ $^d$ & $Ncol$ $^e$\\
     & (GHz) $^a$     & freq. (GHz) $^b$ & opacity (\kms) & (\%) & ($10^{12}$~cm$^{-2}$\,km$^{-1}$\,s) & ($10^{12}$\,cm$^{-2}$)\\
\hline

HCO$^+$ $J$=2--1       & 178.3750563 & 105.8819 & -- & 0.75    & 1.99 & -- \\
H$^{13}$CO$^+$ $J$=2--1 & 173.5067003 & 102.9921 & 1.000(0.098) & 0.29 & 2.00   & 2.0 (0.2) \\
HC$^{18}$O$^+$ $J$=2--1 & 170.3226261 & 101.1021 & 0.459(0.055) & 0.63 & 2.02 & 0.9 (0.1) \\
HC$^{17}$O$^+$ $J$=2--1 & 174.1131691 & 103.3521 & -- & 0.40    & 2.00 & -- \\

HCN $J$=2--1           & 177.2611112 & 105.2207 $^f$ & -- & 0.83    & 3.39 & -- \\
H$^{13}$CN $J$=2--1     & 172.6778512 & 102.5001 $^f$ & 0.842(0.064) & 0.42 & 3.41 & 2.9 (0.2) \\
HC$^{15}$N $J$=2--1     & 172.1079570 & 102.1618 & 0.276(0.035) & 0.41      & 3.43 & 0.9 (0.1) \\

CS $J$=3--2            & 146.9690287 & 87.2396 & 2.471(0.225) & 0.44 & 17.6 & 44 (4.0) \\
C$^{34}$S $J$=3--2      & 144.6171007 & 85.8435 & 0.302(0.047) & 0.31 & 17.5 & 5.3 (0.8) \\

H$_2$S $J_{K_aK_c}$=1$_{10}$--1$_{01}$       & 168.762.7624 & 100.1762 & 1.691(0.157) & 0.68 & 31.7 & 54 (5.0) $^g$ \\
H$_2$$^{34}$S $J_{K_aK_c}$=1$_{10}$--1$_{01}$ & 167.9105160 & 99.6703 & 0.187(0.037) & 0.73   & 31.6 & 5.9 (1.2) $^g$ \\

C$_2$H $N$=2--1 & 174.6915477 & 103.6954 $^f$ & 6.45 (0.04) & 0.32 & 52.1 &  340 (2)  \\

\hline
\end{tabular}
\end{center}
\mbox{\,} \vskip -.5cm
$(a)$ Rest frequencies are taken from the Cologne Database for Molecular Spectroscopy \citep{mul01};
$(b)$ adopting $z$=0.68466, heliocentric frame \citep{wik95}; 
$(c)$ \% of total continuum level;
$(d)$ Coefficients for calculating column density from integrated opacity: $Ncol = \alpha \times \int \tau dv$, assuming purely radiative excitation with CMB photons, i.e., rotation temperature of 4.6~K;
$(e)$ Total column densities, calculated as explained in (d);  
$(f)$ Weighted mean frequency of the hyperfine components;
$(g)$ Total column density assuming an ortho/para ratio of 3.
\end{sidewaystable*}

\end{appendix}

\end{document}